\documentstyle[preprint,revtex]{aps}
\begin{document}
\draft
\begin{title}
\bf{REUNION OF VICIOUS WALKERS:\\ RESULTS FROM
$\epsilon$-EXPANSION}
\end{title}
\author{Sutapa Mukherji\cite{eml1}
and Somendra M. Bhattacharjee\cite{eml2}}
\begin{instit}
Institute of Physics,Bhubaneswar 751 005,India
\end{instit}
\begin{abstract}
The anomalous exponent, $\eta_{p}$, for the decay of the
reunion probability of $p$ vicious walkers, each of length
$N$, in $d$ $(=2-\epsilon)$ dimensions, is shown to come
from the multiplicative renormalization constant of a $p$
directed polymer partition function.  Using renormalization
group(RG) we evaluate $\eta_{p}$ to
$O(\epsilon^2)$.
The survival probability exponent is $\eta_{p}/2$. For
$p=2$, our RG is exact and $\eta_p$ stops at $O(\epsilon)$.
For $d=2$, the log corrections are also determined.
The number of walkers that are sure to reunite is 2 and has
no $\epsilon$ expansion.
\end{abstract}
\pacs{***}
\narrowtext
A question of perennial interest is this: If $p$ vicious
(i.e. mutually avoiding) random walkers in $d$ dimensions
start from a point in space at time $z=0$, how does the
probability of reunion at time $z=N$ decay with $N$?  A
complementary question is the decay of the survival
probability.  Generally, for large $N$, a power law form is
expected signifying certain universality in the behaviour
which we explore in this paper.

One of the main motivations to study  vicious
walkers is to understand and predict the nature of the
phase transitions for interface wetting phenomena, melting
of the commensurate phase, commensurate - incommensurate
(CI) transitions etc, all of which
involve walls or string like objects~\cite{pt,hf}. These walls,
which are statistically
parallel to the $z$ axis, with fluctuations in the
transverse direction may reunite if
defects are present in the system.  The loops formed thereby
are often relevant for
criticality~\cite{hf,smbc,frst,nag,ranis}.  They are also
important in the context of abelian sand pile model of self organized
criticality.~\cite{soc}
These directed strings  can be embedded in
arbitrary $d+1$ dimensions with $d$ as the transverse
spatial dimension.  These are called {\it directed polymers}
(DP).  Note that the preferred direction of the DP plays
the role of time for the random walker.

In spite of its enormous applications, exact results (e.g.,
the decay exponents)
available for reunion of a system of vicious random walkers
are restricted to $d=1$ dimension only, with an approach
for higher dimensions remaining as a challenge.~\cite{pt,hf}
It is clear that the constraints
for the walkers to be reunited at one or both the ends and
their repulsion do change the exponents drastically.  Our
focus is on these exponents for $d>1$.

In this paper, we adopt the DP description. The
viciousness becomes an equal time mutual repulsion for the
DPs.
The reunion and the survival probablities then come
from appropriate partition functions for these interacting
DPs.  Since we are interested in the
asymptotic behaviour, i.e., in the large chain length limit,
we adopt a continuum model.  The merit of the path
integral technique easily enables us to formulate the model
and obtain the exponents of the chain length for the
partition functions in arbitrary $d+1$ dimensions through a
renormalization group (RG) approach.  In this
formulation, the dimensionless hamiltonian for $p$ DPs in
$d+1$ dimensions, each of length $N$, is given by (Note:
$z$ is the preferred direction)~\cite{ns,rsb,smbv}
\begin{equation}
H_p= \frac{1}{2}\sum\limits_{i=1}^p\int\limits_0^N dz
\ {\dot {\bf r}}_i^2 +
v_0\ \sum\limits_{i > j}
\int\limits_0^N dz
\delta^d ({\bf r}_{ij}(z))
\label{eq:hn}
\end{equation}
where ${\dot {\bf r}}_i=\partial {{\bf r}}_i/\partial z$, and
${\bf{r}}_{ij}(z) = {\bf r}_{i}(z)- {\bf r}_{j}(z),\
{\bf r}_i (z) $ being the $d$ dimensional coordiante of a
point at $z$ on the contour of the $i$th chain.  The first
term is the elastic energy, taking care of the connectivity
of the chains.  The viciousness of the walkers is simulated
in the second term by a $\delta$-function interaction at
the same $z$ coordinate with $v_0 > 0$.
Dimensional analysis
shows that $v_0$ is dimensionless and hence
marginal at $d=2$ (uppercritical dimension).
The fact that $d=2$ is special will be reflected
in the later discussion. Our method can be
extended to many body interactions - a case to be discussed
elsewhere~\cite{ranipre}.

The reunion probability for $p$ chains follows from the
constrained partition function $Z_{R,p}$ for DPs with the
restriction that the chains are all tied together at both
the ends at origin. A computationally easier quantity is
the no-reunion or survival probability.  This comes from
the partition function, $Z_{S,p}$, with the restriction
that the walkers are all tied together at the origin
intially, i.e., at $z=0$ but are free at the other end.
These and the exponents are defined generically below ($g = R$
or $S$) as
\begin{equation}
Z_{g,p} = \int {\cal D}{{\bf r}}\ F_g\ e^{-H_p}\
\sim N^{-\psi_{g,p}}. \label{eq:zpr}
\end{equation}
Here, ${\cal D}{\bf r}$ stands for the sum over all paths,
and $F_g$ denotes the end point constraint, implemented by
a $\delta^{d}({\bf r}_i (0)) $ for one end of each chain
($g=R, S$) and a $\delta^{d}({\bf r}_i (N)-{\bf r})$ for the other
end of each chain for reunion at ${\bf r}$.
One can
consider a more general reunion problem where the walkers
can meet anywhere in space at $z=N$.  This, ${\cal Z}_{R,p}$,
requires an
extra integration of the partition function of the
Eq. ~\ref{eq:zpr} type over the end point coordinate, and ${\cal
Z}_{R,p} \sim N^{-\Psi_{g,p}}$.
We have indeed checked explicitly for a few cases that,
the anomalous part of $\psi_{R,p}$ and $\Psi_{R,p}$ are
the same. Incidentally, these exponents have recently been used
to discuss a novel crossover in the Ising model ~\cite{ranis}.

Once the exponents are known, one can address the question
of the critical number of vicious walkers that are sure to
reunite in $d$ dimensions. Conventional analysis [see,
e.g.,~\cite{pt}] shows that this number is determined by
the convergence criterion of the conditional probability
${\cal Z}_{R,p}/Z_{S,p}\sim N^{-\chi}$,
[$\chi=\Psi_{R,p}-\psi_{S,p} $], i.e., whether $\int dN
N^{-\chi}$ is convergent or not.  The critical number is
then obtained from $\Psi_{R,p}-\psi_{S,p} =1$. We
call this number $p_c$.

Our approach is to start with a perturbation expansion
(digrammatically) in the coupling constant, and use the
renormalization group (RG)
approach to go beyond the validity of the
perturbation series~\cite{rsb,smbv}.
The series stumbles on divergences that require standard
dimensional regularization to identify the poles.  The
removal of the divergences demands a renormalization of the
interaction parameter and an overall multiplicative
renormalization constant~\cite{amit}.    This constant
gives rise to an anomalous exponent of the length scale
which is unexpected from a conventional dimensional
analysis. Such constants were
not needed for virial coefficient or other thermodynamic
properties of DPs ~\cite{rsb,smbv} and hence no anomalous
dimension ever appeared there.

We, in this paper, first study
the two chain reunion problem through an exact RG~\cite{rbd}.  It is
then extended to the $p$ chain case. We also study in a
similar way the no-reunion case, without giving much
details.  The results then produce a surprising
``super"universality for $p_c$.  Our procedure is general enough
to answer many other questions than studied here.  These will be
discussed elsewhere.

Had the walkers been noninteracting $(v_0 = 0)$, the
exponents would follow trivially from the normalized
distribution $G({\bf r}\mid z) = (2 \pi z)^{-d/2} \exp
(-r^2/2z)$, for one walk of length $z$ and the end to end
distance ${\bf r}$.  The ``gaussian" exponents
are
\begin{equation}
\psi_{S,p} = 0, \
\psi_{R,p} = p d/2, \ {\rm and} \
\Psi_{R,p} = (p-1) d/2.
\label{eq:expg}
\end{equation}
These gaussian exponents differ from the
exact $d=1$ results for interacting walkers~\cite{pt,hf}
\begin{equation}
\psi_{S,p} = \frac{p(p-1)}{4}, \
\psi_{R,p} = \frac{p^2}{2},\ {\rm and} \
\Psi_{R,p} = \frac{p^2 - 1}{2}.\label{eq:exp1}
\end{equation}
The difference between any of these and the corresponding
gaussian exponent is the anomalous dimension to be denoted
by $\eta_{S,p}$ (for survival) and
$\eta_p$ (for reunion).  A consequence of
Eqs.~\ref{eq:expg} and ~\ref{eq:exp1} is that,
for $d=1$, $p_c=2$ for vicious walkers in contrast to $3$
$(1+2/d$ for general $d$)
for noninteracting walkers.

One might expect $p_c =2$ for $d\leq 2$ on a simple argument.
For two vicious walkers, one can consider
the relative walker who starts from one of
the nearest neighbours of the origin (NNO) but avoids the
origin.  Since for $d=2$ a
random walk is recurrent, one might, ignoring the origin
avoidance,  expect the walker eventually to
come to one of the NNOs again, forcing a reunion.  Our analysis
shows that this is indeed the right answer.

{\it Two chain case:} The simplest case that
can be calculated exactly is
the reunion problem of two walkers.
The partition function which restricts the
two walkers to be tied at the ends at spatial
coordinate ${\bf r} = 0$ can be calculated with
perturbation in the interaction parameter about the free
walkers.  The diagrams representing the terms of the
perturbation series are, e.g., the first three diagrams of Fig
1a with the left most line omitted.  The general term
corresponding to DPs with arbitrary number of
encounters can be calculated easily using the identity for
$m$ gaussian propagators
\begin{equation}
G^m({\bf r}\mid z) = ( 2 \pi z)^{-(m-1)d/2} m^{-d/2} G({\bf
r}\mid z/m).\label{eq:gm}
\end{equation}

In terms of the dimensionless coupling constant $u_0 = v_0
L^{\epsilon}$, ($\epsilon = 2 - d$ and $L$ an arbitrary
length scale) the series for the partition function
is
\begin{equation}
{(2\pi N)^{d}}Z_{R,2} =  1 +
\sum_{n=1}^{\infty} (-u_0/4\pi)^n (4 \pi
NL^{-2})^{n\epsilon/2} {\cal G}_{n+1}(\epsilon/2). \label{eq:z2}
\end{equation}
where ${\cal G}_{n}(x)=\Gamma^{n}(x)/\Gamma(n x)$.
This series, clearly showing divergences at $\epsilon=0$,
requires
renormalization through absorption of these poles.
We quote below, for convenience, the formula for renormalized
$u$ and the corresponding $\beta$-function from Ref. ~\cite{rsb}
\begin{equation}
u_0 = u [ 1 - u/(2\pi \epsilon)]^{-1},\ {\rm and} \
\beta(u) = u \epsilon \ [ 1 - u/(2\pi \epsilon)] .
\label{eq:ur}
\end{equation}
with $u = u^*=2\pi
\epsilon$ as the fixed point where $\beta (u) = 0$.
This interaction renormalization is not sufficient to absorb
even the divergence in
the first order term in $u_0$ of Eq.~\ref{eq:z2}.
Let us define the renormalized partition function as
\begin{equation}
Z_{R,p}{{\mid}}_{r} = R_{R,p}(u)\  Z_{R,p}\ {\rm with}
R_{R,p}(u)=1+\sum_{n\rangle 0}b_n u^n,
\label{eq:renz}
\end{equation}
as the overall
multiplicative renormalization constant $(p=2$).
The divergences can be absorbed completely provided $b_1=(\pi
\epsilon)^{-1},$ and $b_2$ = $3/(4 \pi^2\epsilon^2)$.
In fact, it is possible to sum the whole series to get
$R_{R,2}(u)=[1-u/(2\pi\epsilon)]^{-2}$.

An appeal to the renormalization group equation~\cite{amit} immediately
shows that $\psi_{R,2}$ is
different from the naive engineering dimension $(=d)$,
Eq.~\ref{eq:expg} by $\eta_2 = \gamma_{R,2}(u^*)$ obtained from
\begin{equation}
2 \gamma_{R,p}(u) = \beta(u) \frac{\partial}{\partial u}
\ln R_{R,p}(u) \label{eq:gm2}
\end{equation}
with $p=2$.  We, therefore, have
\begin{equation}
\eta_{2}= \epsilon, \ \psi_{R,2}=2 \ {\rm and} \
\Psi_{R,2}=2 - d/2.
\label{eq:exp2}
\end{equation}
in agreement with the $d=1$
exact results of Eq. \ref{eq:exp1}~\cite{rbd}.

{\it $p$ chains:} Let us first consider the partition
function $Z_{R,p}$ for $p$ chains tied together at the
origin at both the ends.  The terms in the perturbation
expansion upto second order in $v_0$ correspond to the
diagrams shown in Fig. 1a.  The zeroth order gaussian term
(Fig. 1a.1) contributes $(2\pi N)^{-pd/2}$.

The second order diagrams are of several types: ladder type
involving two chains (Fig. 1a.3), three chains connected by
interaction (Fig. 1a.4), and two pairs connected
independently (Fig.  1a.5). The first and the last of these
and the first order diagram (Fig.1a.2) are already
evaluated in Eq.  ~\ref{eq:z2}, since these are just two
chain terms.  The most crucial contribution is from
Fig. 1a.4 which {\it connects three chains}.  The two chain
terms are important for renormalization, but
the $O(\epsilon^2)$ contribution to the exponents, ultimately,
comes from this diagram.

The contribution of Fig. 1a.4 involves
$(2 \pi N)^{-(p-3)d/2}$ for the noninteracting
$(p-3)$ chains. To evaluate the connected piece, one
has to use an identity similar to Eq.~\ref{eq:gm} for $m$
gaussian propagators, and the markovian and normalization
properties of $G$.  Details will be given
elsewhere~\cite{ranipre}.
The final expression is
\begin{eqnarray*}
{\displaystyle{{(2 \pi N)^{-pd/2}}{N^{\epsilon}
{(4\pi)}^{-d}}}} & [ &
{\cal G}_3(\hat{\epsilon}) \ {}_2F_1
\left ( \hat{\epsilon},\hat{\epsilon};
3\hat{\epsilon}; 3/4\right )
+ \nonumber\\ & & \left (3/4
\right )^{\hat{\epsilon}}
\Gamma(-\hat{\epsilon})\Gamma^{-1}(1-\hat{\epsilon})
{\cal G}_2(\epsilon)
\ {}_3F_2\left ( \epsilon,\epsilon,1;2\epsilon,1+
\hat{\epsilon};
3/4\right )\ ],
\label{eq:zr3c}
\end{eqnarray*}
where $\hat{\epsilon} = \epsilon /2$, ${}_3F_2$ is the generalized
hypergeometric function, and ${\cal G}_n(x)$ is defined after
Eq. \ref{eq:z2}.  Combining all the terms with
appropriate symmetry factors (see Fig. 1c) and expanding
each one around $\epsilon = 0$, we find that
\begin{equation}
{\displaystyle{\frac{Z_{R,p}}{(2\pi N)^{-pd/2}}}}  =  1 -
\frac{u_0}{2\pi\epsilon}\ (_2^p) (2 + \epsilon \ln x)
+ \frac{u_0^2}{4\pi^2\epsilon^2} (C - \epsilon D
+\epsilon C\ln x) + ...\ ,
\label{eq:zpf}
\end{equation}
where $C = (_2^p)(p^2 - p +1)$, $D=3(_3^p)\ln (3/4)$,
and $x = 4\pi NL^{-2}$.

Substituting $u$, from Eq. \ref{eq:ur}, for $u_0$ (note that $u_0$
still pertains to two body $\delta$ function interaction)
in $Z_{R,p}$, Eq.~\ref{eq:zpf}, one sees
that the series still requires an
overall multiplicative renormalization constant $R_{R,p}(u)
= 1 + b_1 u + b_2 u^2 +...$ (see Eq.~\ref{eq:renz}).  Minimal
subtraction of the poles~\cite{amit} yields
$b_1 = (_2^p)/(\pi \epsilon)$, and
$ b_2 =  (C + \epsilon \ D)/(2\pi\epsilon)^{2}$.

The anomalous exponent comes from $\gamma_{R,p}(u)$ of
Eq.~\ref{eq:gm2}.  Using
the $\beta$-function, Eq.~\ref{eq:ur}, we get
\begin{equation}
2\gamma_{R,p} (u) = \epsilon u
[b_1 + ( 2 b_2 - b_1 (2\pi\epsilon)^{-1} - b_1^2) u + ..].
\label{eq:gmu}
\end{equation}

Finiteness of $\gamma_{R,p}(u)$ in the limit $\epsilon
\rightarrow 0$ is guaranteed by the exact cancellation of
the $O(\epsilon^{-2})$ term in $2 b_2 - b_1
(2\pi\epsilon)^{-1} - b_1^2$ leaving alone the
$O(\epsilon^{-1})$ term. In fact, the finiteness criterion
of $\gamma_{R,p} (u)$ as $\epsilon
\rightarrow 0$ dictates the cancellation
of all but the $O(\epsilon^{-1})$ part in all higher orders
of $u$. Therefore, to calculate the $O(\epsilon^2)$ term in
$\gamma(u^*)$, no higher order term in the renormalization
constant in $R_{R,p}(u)$ is required. This $O(\epsilon^2)$
term can be traced back to the three chain connected
diagram.

We finally obtain from Eq. \ref{eq:gmu} the anomalous exponent
for reunion as
\begin{equation}
\eta_p \equiv \gamma_{R,p}(u^*) = \ (_2^p)\
\epsilon + 3 \ (_3^p) \ \ln
(3/4)\ \epsilon^2 + O(\epsilon^3).\label{eq:re}
\end{equation}

The survival case, can also be considered in an identical
fashion. The diagrams are shown in Fig. 2b.
The unexpected result we find is that, upto
$O(u^2)$, the renormalization constant $R_{R,p}(u)$ is the
square of the renormalization constant needed for
$Z_{S,p}$. We have seen it exactly for $p=2$.  Though it
seems rather obvious or intuitive~\cite{dup2}, still
a general proof is lacking.  We conclude
\begin{equation}
\psi_{S,p} = \eta_{S,p}
=\eta_p/2.\label{eq:nr}
\end{equation}
This is a scaling relation not hitherto recognized at all.

At the upper critical dimension $d=2$, the $\beta$-function
has only the trivial fixed point.  Its integration gives $
u(L) = u_0 [1 + (u_0/2\pi) \ln (L/L_0)]^{-1}$ where $u_0 =
u(L_0)$.  The renormalization group equation, furthermore,
via the method of characteristics~\cite{amit}, leads to the log
correction for the reunion probability, as
\begin{equation}
Z_{R,p} \sim N^{-(p-1)} \ [u_0 \ln (N/N_0) ]^{-p(p-1)/2}
\nonumber
\end{equation}
where $N_0=L_0^2$. Similarly, the survival probability,
instead of remaining $N$ independent as for the gaussian
case, now has a slow deacy as $[\ln (N/N_0)]^{-p(p-1)/4}$.

To determine $p_c$, we now solve
$(p-1)d+\eta_p=2$ in an iterative way. Upto $O(\epsilon^2)$,
$p_c=2$, even after incluidng the log corrections at $d=2$.
Taking into account the exact
result ~\cite{hf} for $d=1$, we predict that $p_c=2$
for $d<2$ and is independent of $d$.
Of course, two walkers
are not certain to reunite for $d>2$.
Still, the fact that the interaction can make the
critical number ``super"universal (at least weakly, i.e.,
for $d\leq 2$), is surprising.

In summary, we emphasize that for the problem of vicious
walkers, that can be visualized via the path integrals as
directed polymers, an $\epsilon=2-d$ expansion is sucessful
in obtaining the essential results.  The key feature is the
evaluation of the overall multiplicative renormalization
constants for the partition functions.  The relevant exponents
for $p$ chains are computed to $O(\epsilon^2)$.
Extensions to still higher orders in $\epsilon$ seem not out of
reach. Our analysis
shows that for $O(\epsilon^2)$, three chains should be
collectively aware of each other. Such collective features are
expected to continue in higher orders also.  For $p=2$, we
obtain exact $O(\epsilon)$
results.  A significant result is
the scaling relation Eq.~\ref{eq:nr} between the anomalous
exponents for reunion and survival probabilities.  The
speciality of the dimension $d=2$ shows up through the
dependence of the probabilities on $\ln N$ with $p$ dependent
powers.  Furthermore,
though 3 (2) noninteracting walkers are sure to unite in
one (two) dimensions, we find the number is 2 for vicious
walkers, independent of $d<2$.  There seems to be a
superuninversality about this number.  Numerical
verification of these predictions, especially on fractals,
would be welcome.

A major part of this work was done at ICTP, Trieste, during
our visit there under the Federation scheme.  We thank ICTP
for its warm hospitality.

\figure{Diagrammatic expansion upto $v_0^2$ for $p\geq 3$
chains tied at (a) both ends and (b) one end.  The solid lines
represent the
chains and the dashed lines represent the interaction at
same chain length $z$ (the involved points on the chains
have the same spatial coordinates.). Only a few chains are
shown.  Rules for evaluation:
(i) a factor of $G$ for each piece of the solid lines, (ii)
$-v_0$ for each dotted line, (iii) an integration over the
coordinates of the interaction points, and (iv) a time
ordered integration of the $z$ coordinates. See Ref.
{}~\cite{rsb,smbv}.  (c)
Symmetry factors.  These numbers take into account the
number of diagrams with identical contribution but
generated either through a mere permutation of the chains
or by reversal of time ordering.}
\end{document}